\documentclass[aps,pre,twocolumn,amsmath,10pt,groupedaddress,superscriptaddress,amssymb]{revtex4-1}

\usepackage{graphicx}   
\usepackage{subfigure}  
\usepackage{hyperref}  

\newcommand{\dd}{\mathrm{d}}

\newcommand{\s}{\sigma}
\newcommand{\be}{\begin{equation}}
\newcommand{\ee}{\end{equation}}

\DeclareMathOperator*{\covar}{Covar}
\DeclareMathOperator*{\var}{Var}

\DeclareMathOperator*{\sign}{sign}

\DeclareMathOperator*{\atanh}{atanh}

\begin{document}

\title{Anomalous finite size corrections in random field models}

\author{C. Lucibello}
\affiliation{Dipartimento di Fisica,``Sapienza'' University of Rome, P.le A. Moro 2, I-00185, Rome, Italy}
\author{F. Morone}
\affiliation{Levich Institute and Physics Department, City College of New York, New York, New York 10031, USA}
\author{G. Parisi}
\affiliation{Dipartimento di Fisica, IPCF-CNR, UOS Roma, and INFN, Sezione di Roma1, ``Sapienza'' University of Rome, P.le A. Moro 2, I-00185, Rome, Italy}
\author{F. Ricci-Tersenghi}
\affiliation{Dipartimento di Fisica, IPCF-CNR, UOS Roma, and INFN, Sezione di Roma1, ``Sapienza'' University of Rome, P.le A. Moro 2, I-00185, Rome, Italy}
\author{Tommaso Rizzo}
\affiliation{IPCF-CNR, UOS Roma, ``Sapienza'' University of Rome, P.le A. Moro 2, I-00185, Rome, Italy}

\begin{abstract}
The presence of a random magnetic field  in ferromagnetic systems leads, in the broken phase, to an anomalous $O(\sqrt{1/N})$ convergence of some thermodynamic quantities to their asymptotic limits. Here we show a general method, based on the replica trick, to compute analytically the $O(\sqrt{1/N})$ finite size correction to the average free energy. We apply this method to two mean field Ising models, fully connected and random regular graphs,  and compare the results to exact numerical algorithms. We argue that this behaviour is present in finite dimensional models as well.
\end{abstract}

\maketitle

\section{Introduction}
\label{sec:intro}
There is a large interest in understanding the behaviour of ferromagnetic
systems in random magnetic fields. Also the most simple model one can conceive,  the Random Field Ising Model (RFIM), has a plethora of applications to condensed matters problems \cite{Fishman1979, Dagotto2005, Vink2006}, while a complete understanding  of its static and dynamical properties is still lacking. Only in recent years some long-standing issues such as the presence of a glassy phase \cite{Krzakala2010, Krzakala2010a} and the universality of the zero-temperature fixed point \cite{Fytas2013} have been partially settled.

Here we contribute to the literature on such random systems discussing the convergence properties of the average free energy  to its thermodynamic limit. Finite size corrections
to the free energy have been investigated by the authors in two papers \cite{LucibelloER,LucibelloRRG} focusing on the paramagnetic phase of disorder Ising models on diluted graphs, in an effort to characterize the free energy correction induced by loops to the Bethe mean field approximation.

Perturbative expansions around mean field approximations are not yet numerically and analytically manageable, in spite of the large
amount of work done in these years \cite{Montanari2005, Chertkov2006, Parisi2006}.  It was suggested in Refs. \cite{LucibelloER,LucibelloRRG} that the
study of finite size correction for infinite range systems has many
points in common with the computation of loop corrections in finite
dimensional problems. In order to make progresses the case of finite
size correction to the average free energy was carefully studied in the case of Erd\"{o}s-R\'enyi and random regular graphs \cite{LucibelloER,LucibelloRRG}. The correction were computed in the high temperature phase and they were divergent at the transition point. We show in this paper that the $O(1/N)$ corrections are overshadowed in the whole low temperature
phase, because $O(\sqrt{1/N})$ corrections arise. The computation of these
corrections is the goal of this paper. We present both numerical
simulations and analytical computations to support this claim.

These $O(\sqrt{1/N})$ corrections do not arise from loops, but they come
from disorder induced fluctuations in the weight of the two
macroscopically different states.  A more complex version of this
phenomenon should take place in the presence of many different
equilibrium states. This phenomenon is also present
 in finite dimensional models, as it is suggested by some preliminary numerical we conducted. From the technical point of view the
analytical computations in the replica framework are connected to the
existence of multiples solutions of the mean field equations \cite{Dotsenko1997}\cite{Dotsenko2006}.

In the following Section we present the general analytical framework that can be employed to compute the $O(1/\sqrt{N})$ corrections to the average free energy density. Notice that the same corrections are of order $O(\sqrt{N})$ when referred to extensive quantities, therefore the indistinct use in the text of  $O(1/\sqrt{N})$ and $O(\sqrt{N})$ should not confuse the reader. In Sections \ref{sec:fullyconn}  and \ref{sec:rrg} we apply this formalism to the fully connected and to a diluted  Ising model respectively, checking numerically the consistency of the results.

\section{General Formalism}
\label{sec:formalism}

We consider for concreteness a system of $N$ Ising spins, $\s_i=\pm 1$, and external random fields. The following arguments though apply  to a general class of models possessing  $O(m)$ symmetry (once the average over disorder is taken). On a given graph $G$, the Hamiltonian of the RFIM is
\begin{equation}
\mathcal{H}=-J \sum_{(i,j)} \s_i \s_j - \sum^N_{i=1} h_i\, \s_i,
\label{hamiltonian}
\end{equation}
where the first sum is over adjacent spins, $J\geq 0$ is a ferromagnetic coupling and the the fields $h_i$ are quenched i.i.d. random variables with zero mean and unit variance, i.e. $\mathbb{E}[h_i]=0$ and $\mathbb{E}[h_i\,h_j]=\delta_{ij}$. 

It takes a simple argument to show that, at least at zero temperature and for $J$ large enough, the subleading term in $N$ to the average energy $E(N)$ is of order $O(\sqrt{N})$. In fact in this case, for a given realization of the external fields $\{h_i\}$, the Gibbs measure is concentrated on the configuration with minimum energy, the candidates being the one  with all the spins up and  the one with all the spins down. For a given graph with $M=O(N)$ edges, the energies of the two states, let us call them $E_+$ and are $E_-$ respectively, are given by
\begin{equation}
E_{\pm} = - M J\mp \sum^N_{i=1} h_i  \qquad  \text{ for } J\gg 1 .
\end{equation}
The sum in the r.h.s. is a random variable of variance $N$, converging to a Gaussian variable in the thermodynamic limit. Therefore it is easy to see that for the  average energy $E(N)=\mathbb{E}[\min(E_+,E_-)]$ we have
\begin{equation}
\begin{aligned}
E(N)=
 - M J  -\sqrt{\frac{2}{\pi}}\sqrt{N}+o\big(\sqrt{N}\big)  \qquad  \text{ for } J\gg 1 .
\end{aligned}
\end{equation}
It turns out that the $\sqrt{N}$ subleading behaviour we found in this limit case is present in the whole ferromagnetic region in the $J-T$ (coupling-temperature) plane. 

In fact in the ferromagnetic phase the statistical weight is concentrated on two disconnected regions of the configuration space, having positive and negative magnetization respectively, separated by a free energy barrier exponentially increasing in $N$. While the two regions are completely equivalent in the pure ferromagnetic system, once the disordered external field is turned on their free energies start to differ of a random quantity of order $O(\sqrt{N})$. The total free energy of the system is then  simply given by the minimum among the two, except for some exponentially decaying terms.
Let us first define, for a given realization of the disorder, the free-energies $F_+$ and $F_-$ as the ones corresponding to configurations having positive and negative magnetization respectively (for simplicity we assume $N$ to be odd). We assume, and verify a posteriori for the RFIM, that in the ferromagnetic phase the difference between the free energies of the two states, $F_+-F_-$, is of order $O(\sqrt{N})$.  Then the average free energy of the system is given by
\begin{equation}
\begin{aligned}
F(N)& = -\frac{1}{\beta}\,\mathbb{E}\left[ \log\left(e^{-\beta F_+}+e^{-\beta F_-}\right)\right]\\
&= \mathbb{E}\big[\min(F_+,F_-)\big]+\textit{exp. vanish. terms}.
\end{aligned}
\end{equation}

From these premises it follows simply that the average free energy has an expansion in $N$ of the form
\begin{equation}
F(N)=  f_0 N +f_1\sqrt{N} +o(\sqrt{N}),
\label{FN}
\end{equation}
where the coefficients $f_0$ and $f_1$ of the expansion are defined by
\begin{equation}
f_0=\lim_{N\to+\infty}\frac{1}{N}\, \mathbb{E}\big[\,F_+\,\big]=\lim_{N\to+\infty}\frac{1}{N}\, \mathbb{E}\big[\,F_-\,\big],
\label{f0}
\end{equation}
\begin{equation}
f_1=\lim_{N\to+\infty}\frac{-1}{\sqrt{N}}\, \frac{\mathbb{E}\big[\,|F_+ - F_-|\,\big]}{2}.
\label{f1}
\end{equation}
To compute $f_1$ we extend  a method developed in Refs. \cite{Aspelmeier2003a} and \cite{Parisi2008,Parisi2009,Parisi2010a}. At fixed $N$ and for a given realization of the disorder, let us call $Z_+$ and $Z_-$, the partition function constrained to the configurations with positive and negative magnetization. Considering $n$ and $m$ replicated systems respectively, one can easily see that for small $n$ and $m$ we have 
\begin{equation}
\begin{aligned}
-\frac{1}{\beta}\log\,\mathbb{E}\big[\,Z_+^{n}\,& Z_-^{m}\big]\sim\, n\,  \mathbb E [F_+] + m\,  \mathbb E [F_-]-\frac{n^2}{2} \beta\var(F_+)\\
&  -\frac{m^2}{2} \beta\var(F_-) -n m\beta\covar(F_+,F_-).
\end{aligned}
\label{mnexp}
\end{equation}
Obviously, for symmetry reasons, $\mathbb E [F_+]=\mathbb E [F_-]$ and $\var [F_+]=\var [F_-]$.
The free energies $F_+$ and $F_-$ are jointly distributed random variables. If the limit
\begin{equation}
\begin{aligned}
\Delta^2 &\equiv\lim_{N\to\infty}\frac{1}{2 N}\big[\var(F_{+}) -\covar(F_+,F_-) \big]\\
&=\lim_{N\to\infty}\frac{1}{4 N}\mathbb{E}\big[(F_{+} - F_-)^2 \big]
\end{aligned}
\label{defDelta}
\end{equation}
is non-zero, the rescaled variables $f_+$ and $f_-$, defined by
\begin{equation}
f_\pm =\frac{F_\pm - f_0 N}{\sqrt{N}},
\end{equation}
become non-trivially jointly distributed Gaussian random variables for large $N$. In fact, while higher orders cumulants in the $n$ and $m$ expansion of Eq. \eqref{mnexp} are  of order $O(N)$, they give vanishing contributions to $f_+$ and $f_-$.

Since asymptotically $\frac{f_+-f_-}{2}$ is itself a Gaussian random variable of zero mean and variance $\Delta^2$, it follows that the coefficient $f_1$  of Eqs. \eqref{FN} and \eqref{f1} is given by 
\begin{equation}
f_1 = -\int \frac{\dd z}{\sqrt{2\pi \Delta^2}}\ |z|\ e^{-\frac{z^2}{2\Delta^2}} =-\Delta\sqrt{\frac{2}{\pi}}.
\label{f1delta}
\end{equation}

The computation of the $n$ and $m$ expansions in Eq. \eqref{mnexp} can be performed using the standard replica techniques.
Since it is impractical to work with the partial partition functions $Z_+$ and $Z_-$, we define the partition function $Z_H$ of the system having an additional deterministic external field $H$ acting on all the spins. Then we define the replicated free energy $\phi(n,m)$  as
\begin{equation}
\phi(n,m) =\lim_{H\to 0^+}\lim_{N\to\infty} -\frac{1}{\beta N}\log\,\mathbb{E}\big[\,Z_H^{n}\, Z_{-H}^{m}\big].
\label{phi}
\end{equation}
This definition is completely consistent with the $l.h.s.$ of Eq. \eqref{mnexp} in the ferromagnetic region, and is even more physically sound in the paramagnetic region, allowing to treat with a unified formalism the whole phase space.

We compute $\phi(n,m)$ for integer values of $n$ and $m$ using the saddle point technique for $N\to\infty$. We obtain as usual a variational expression depending  on a replicated order parameter and the particular structure of the field in Eq. \eqref{phi} implies that we must look for a solution where the first $n$ replicas are positively magnetized while the remaining $m$ replicas are negatively magnetized. Then we make an analytical continuation of the solution to real value of $n$ and $m$. By definition \eqref{phi} the  $0-th$ order term of the expansion is zero. Since we are interested only in the quadratic terms of the $\phi(n,m)$ expansion, and since the $\phi(n,m)$ is variational in some order parameter, our calculations will involve only the order parameter computed at $n=m=0$ and we do not have to consider its $n$ and $m$ dependence. As always happens with calculations involving replicas,  the exchange of the $n$ and $N$ limits will be done without particular care.

It is interesting to note that the $O(\sqrt{N})$ corrections will be obtained evaluating expression \eqref{mnexp}
 at the leading $O(N)$ order, that is by taking the saddle point, while the more common $O(1)$ corrections are typically controlled by  Gaussian fluctuations around the saddle point.
 
In the following Sections we will explicitly compute $\phi(n,m)$ given by Eq. \eqref{phi} in two mean field models.

\section{Fully connected}
\label{sec:fullyconn}
The first case we consider is that of the RFIM on the fully connected graph. Here the first sum in the Hamiltonian \eqref{hamiltonian} runs over $N(N-1)/2$ edges, and the coupling $J$ has to be rescaled by a factor $N^{-1}$ in order to obtain an extensive thermodynamic behaviour. It is then easy to compute the replicated free energy of the system, defined in Eq. \eqref{phi}, through standard Hubbard-Stratonovich transformation is given by 
\begin{equation}
\begin{aligned}
&\phi_{FC}(n,m) = \min_{x\geq 0} \bigg\{\frac{1}{2} J (n+m)\, x^2 \\
 	&\ -\frac{1}{\beta}\log \mathbb{E}_h\left[ \big(2\cosh\beta(J x+h)\big)^n \big(2\cosh\beta(-J x+h)\big)^m\right]\bigg\}.
\end{aligned}
\label{phiFC}
\end{equation}
Notice that the order parameter $x$, the magnetization of the up-state, enters in the last term of Eq. \eqref{phiFC} both with a plus and a minus sign. 
As a technical note, in Eq. \eqref{phiFC} the minimization condition does not ever turn to a maximization when the number of replicas is small, as it happens in glassy models \cite{Parisi1987},  since the dimension of the order parameter in the full replica space is $n+m$ and is always greater than zero in our calculations.

To obtain the $O(\sqrt{N})$ contribution to the free energy we have to compute the small and $n$ and $m$ expansion of $\phi_{FC}(n,m)$. As already discussed at the end of the previous section, only the $n=m=0$ saddle point solution, given by the non-negative solution of 
\begin{equation}
x=\mathbb{E}_h\tanh \beta (J x + h),
\end{equation}
appears in the second order expansion of the replicated action. Therefore the coefficient $\Delta$, defined in Eq. \eqref{defDelta} and characterizing the $O(\sqrt{N})$ correction though \eqref{f1delta}, is given by
\begin{equation}
\begin{aligned}
\Delta_{FC}^2 =&\, \frac{1}{2\beta^2}\big\{\mathbb{E}_h[(\log\cosh \beta(J x +h))^2]\\
&-\mathbb{E}_h[\log\cosh \beta(J x +h)\log\cosh \beta(J x - h)]\big\}.
\end{aligned}
\label{deltaFC}
\end{equation}
Last expression can be easily computed for any value of $J$ and $\beta$. In order to compare the analytical prediction with exact results, it is easier to consider the zero temperature limit of \eqref{deltaFC}. This is given by
\begin{equation}
\lim_{T\to 0}\Delta_{FC}^2= \frac{1}{4}\,\mathbb{E}_h\big(|J x +h|-|J x -h|\big)^2
\label{deltaFCT0}
\end{equation}
with $x$ non-negative solution of
 \begin{equation}
x=\mathbb{E}_h\sign(J x + h).
\end{equation}

We focus  on the zero temperature limit for two reason: the zero temperature fixed point is the one controlling the flow of the renormalization group also starting from finite temperature \cite{Bray1985}; the ground state of the RFIM can be computed efficiently using exact numerical algorithms.

We implemented  an exact and very efficient algorithm to compute the ground state  of a fully connected RFIM, that takes advantage of the topological equivalence of all the spins.
In fact it is easy to realize that among all the configurations having a certain total magnetization $M=\sum_i \s_i$, the one with lowest energy is the one where only the first $\frac{N-M}{2}$ spins with the lowest external field are down.

\begin{figure}
\includegraphics[width=\columnwidth]{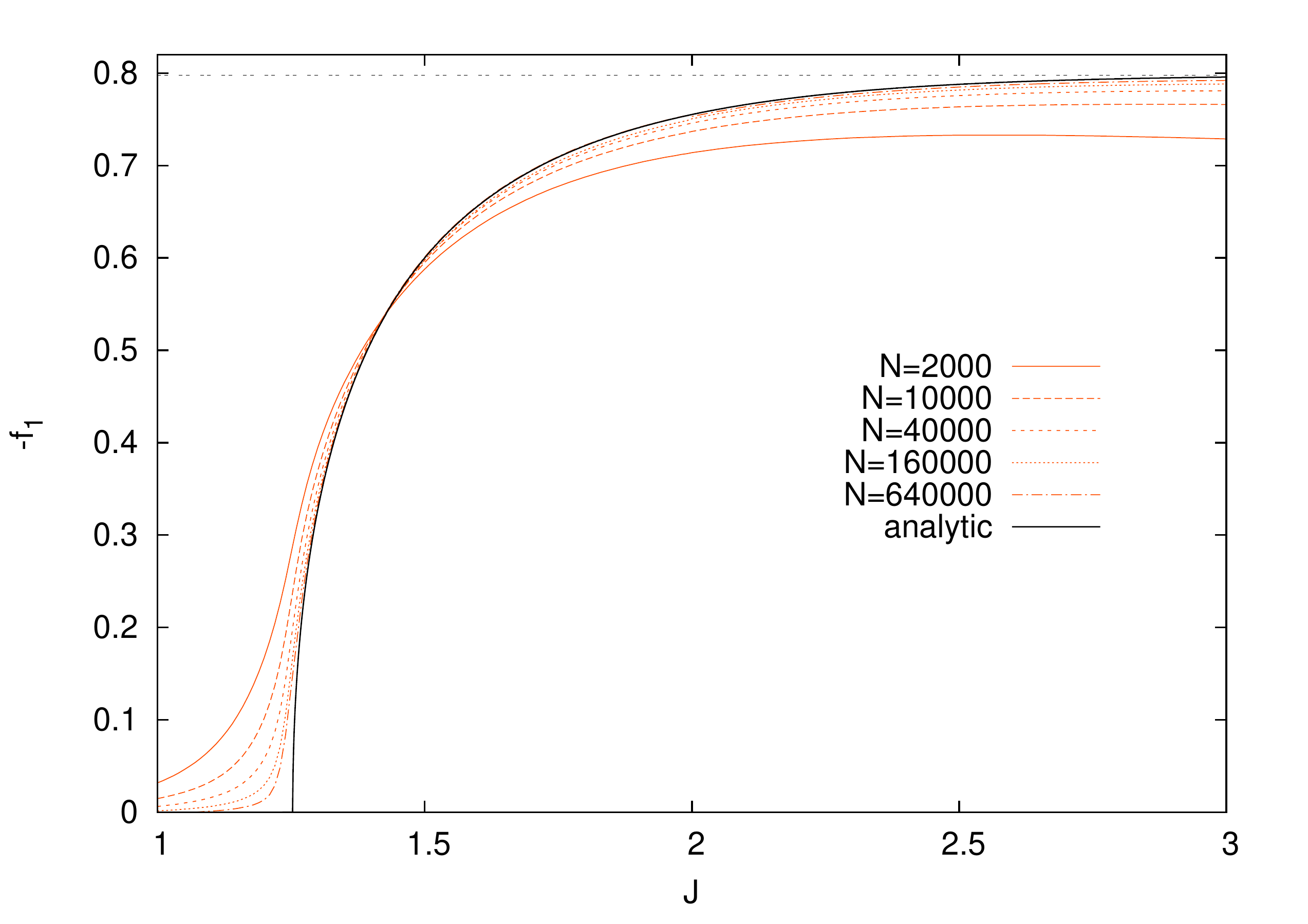}
\caption{The numerical estimates, obtained through the algorithm described in the text, of the coefficient of the $O(\sqrt{N})$ correction to the average free energy in the RFIM at zero temperature on the Fully Connected Graph, for many sizes of the system (\textit{orange lines)}. The numerical curves extrapolate to the analytical prediction of Eqs. \eqref{f1delta} and \eqref{deltaFCT0} for large $N$ (\textit{black line}). The dashed black line is the asymptotic value $\sqrt{\frac{2}{\pi}}$.}
\label{fig:fullyconn}
\end{figure}

Therefore, once the spins are sorted according to their external fields (an operation of time complexity  $\Theta(N\log N)$), we have to look only to these $N+1$ configurations characterized by $M=-N,-N+2,\dots,N$ to find the ground state (an $\Theta(N)$ operation). We have thus produced an algorithm of time complexity $\Theta(N\log N)$ and with $\Theta(N)$ memory requirements. This is a great improvement over the min-cut algorithm that we use on diluted graphs (as we shall explain in next Section), that has time and memory complexity $\Theta(N^3)$  for fully connected graphs.
With this algorithm we where able to perform highly precise averages of systems  up to $\sim 10^6$ spins. We then subtract the leading order (in $N$) term $f_0 N$, that can be computed exactly, to obtain the numerical estimate of the coefficient $f_1$ up to subleading finite size effects.
In Fig. \ref{fig:fullyconn} we show the perfect agreement between the results of our exact algorithm  and the analytical prediction $f_1 =-\Delta\sqrt{\frac{2}{\pi}}$, with $\Delta$ given by Eq. \eqref{deltaFCT0}, for the RFIM at $T=0$.

\section{Random regular graphs}
\label{sec:rrg}
The analytical computation of the $O\big(\sqrt{N}\big)$ correction to the average free energy in diluted models is slightly more involved. Here we focus on the random regular graph (RRG) ensemble. Each element of the ensemble is chosen uniformly at random among all the graphs where each node has fixed connectivity $z$. The RRG and other diluted graph ensembles have the property of being locally tree-like, in the sense that each finite neighborhood of a randomly chosen node is with high probability a tree in the large $N$ limit, and the density of finite loops goes to zero \cite{Montanari2009}. This property allows for the analytical solution  of such models, at least at the leading order in $N$.

Following the lead of Ref. \cite{Parisi2010a}, we apply the replica technique to Eq. \eqref{phi}, to obtain the replicated free energy
\begin{equation}
\begin{aligned}
\phi_{RRG}&(n,m) = \min_{\rho} \Big\{\frac{z}{2}\, \phi_{edge}[\rho] - (z-1)\,  \phi_{site}[\rho]\Big\},
\end{aligned}
\label{phiRRG}
\end{equation}
where 
\begin{equation}
\begin{aligned}
\phi_{edge}&[\rho]=\log\Big[ \mathbb{E}_{h}\sum_{\s_1,\s_2} \rho^{z-1}(\s_1)\,  \\
&\times e^{\beta (h_1\sum_{a}\s_1^a+J\sum_{a}\s^a_1\s^a_2 +h_2\sum_{a}\s^a_2)}\,\rho^{z-1}(\s_2)\Big].
\end{aligned}
\end{equation}
and
\begin{equation}
\phi_{site}[\rho] = \log\Big[\mathbb{E}_h \sum_\s e^{\beta h\,\sum_{a=1}^{n+m} \s^a}\,\rho^z(\s)\Big]
\end{equation}
A similar  expression, for the replicated free energy of spin glasses on RRG, was derived in Ref. \cite{DeDominicis1990} (see also Eq. (7) of Ref. \cite{Rizzo2009} for a more general formulation) and relies on the hypothesis that the graph contains few short loops. 

Eq. \eqref{phiRRG}  is a straightforward generalization of these results, the difference being that  the replicas are divided into two blocks of size $n$ and $m$ respectively.
The minimization  condition in Eq. \eqref{phiRRG} is imposed over all the functions $\rho(\s)=\rho(\s^1,\ldots,\s^{n+m})$ of a $n+m$  replicated Ising spin, taking particular attention to constrain the first $n$ and last $m$ replicas to be in the up and down state respectively.  
Here, as in the fully connected case, the dimension of the order parameter is always positive, since it is $2^{n+m}$. Therefore we do not incur in a common peculiarity of replica calculations, the exchange between minimization and maximization conditions \cite{Parisi1987}.

\begin{widetext}
Under replica symmetry assumptions the order parameter $\rho(\s)$ can be parametrized in the form

\begin{equation}
\rho(\s)=\int \frac{\dd P(u^+,u^-)}{(2\cosh \beta u^+)^n  (2\cosh \beta u^-)^m}\, e^{\beta(u^+\sum_{a=1}^n \s^a +u^-\sum_{a=n+1}^m \s^a)}.
\end{equation}

In the small $n$ and $m$ limit the minimum condition gives 
\begin{equation}
P(u^+,u^-)=\mathbb{E}_h\int \prod^{z-1}_{k=1} \dd P(u_{k}^+,u_{k}^-)\ \delta\bigg(u^+ -g(h+\sum_k u_{k}^+)\bigg)\,\delta\bigg(u^- -g(h+\sum_k u_{k}^-)\bigg),
\label{puu}
\end{equation}
\end{widetext}
where the function $g(x)$  is the usual cavity iteration rule $g(x) =\frac{1}{\beta}\atanh(\tanh(\beta J)\tanh(\beta x))$. The vanishing auxiliary external field $H$  of Eq. \eqref{phi} selects the solution of Eq. \eqref{puu}, supposed to be unique, such that $\overline{u^+}\geq 0 \geq \overline{ u^-}$ (denoting with $\overline{\,\cdot\,}$ the expectation over $P(u^+,u^-)$). The algorithmic equivalent of the distributional equation \eqref{puu} is nothing else than the standard Belief Propagation algorithm applied two times with opposite initializations on the same sample.

Expanding the replicated free energy $\phi_{RRG}(n,m)$ to the second order in $n$ and $m$ we can than derive the $O(\sqrt{N}$) coefficient of the free energy according to Eqs. \eqref{mnexp} and \eqref{f1delta}. We note that only the fixed point distribution $P(u^+,u^-)$ computed at $n=m=0$, that is the solution of \eqref{puu}, contributes to the quadratic order of $\phi(n,m)$.
The coefficient $\Delta$ appearing in $f_1=-\Delta\sqrt{2/\pi}$ than reads
\begin{equation}
\begin{aligned}
\Delta_{RRG}^2 =&\,  \frac{1}{2\beta^2}\bigg\{\frac{z}{2}\big(\mathbb{E}[A^2_+]-\mathbb{E}[A_+ A_-]\big)\\
&-(z-1)\big(\mathbb{E}[B^2_+]-\mathbb{E}[B_+ B_-]\big)\bigg\}.
\end{aligned}
\label{deltarrg}
\end{equation}
The terms $A_{\pm}$ and $B_{\pm}$ stem from the edge and site replicated free energies ($\phi_{site}$ and $\phi_{edge}$) respectively. The expectations $\mathbb{E}[\,\cdot\,]$ are  over both the distribution of the external random field and of the cavity fields.  The site terms $B_+$ and $B_-$ are defined by
\begin{equation}
B_{\pm}=\log \frac{2 \cosh \beta (h +\sum_{k=1}^{z} u_{k}^\pm)}{\prod_{k=1}^{z} 2\,\cosh \beta u_{k}^\pm}.
\end{equation}
Here the fields  $u_{k}^+$  and $u_{k}^-$ are distributed according to $P(u_{k}^+,u_{k}^-)$, solution of Eq. \eqref{puu}, and $h$ is distributed as the external random fields.
The edge terms $A_+$ and $A_-$, appearing in Eq. \eqref{deltarrg}, are defined by
\begin{equation}
A_{\pm}=\log\sum_{\s_1,\s_2} \frac{\exp{\beta( h^{\pm}_1 \s_1 +J\s_1\s_2 + h^{\pm}_2\s_2)}}{\prod_{k=1}^{z-1}4\,\cosh \beta u_{1k}^\pm\,\cosh \beta u_{2k}^\pm}
\end{equation}
The random field $h_1^+$ is distributed as $h+\sum_{k=1}^{z-1} u_{1k}^+$, where $h$ is an external random fields, and analogous definitions follows for the other cavity fields.

We notice that only in the terms $\mathbb{E}[A_+ A_-]$ and $\mathbb{E}[B_+ B_-]$ of Eq. \eqref{deltarrg} the full joint distribution $P(u^+,u^-)$ is needed, not only its marginals.

The computation of $\Delta^2$, and therefore of the analytical finite size correction $f_1$, to a high level of precision through Eq. \eqref{deltarrg}, is a computationally easy task. We solved numerically the fixed point condition Eq. \eqref{puu} through a population dynamic algorithm. In this case each element of the population is a couple of cavity messages, $u^+$ and $u^-$, each of them encountering the same external random fields $h$ during the iterations of the algorithm. As initial condition, in each couple the message  $u^+$ is set to a high positive value, while the message $u^-$ is set to a low negative value. 

In the paramagnetic phase the stable solution of Eq. \eqref{puu} takes the trivial form $P(u^+,u^-)=P(u^+)\delta(u^+ - u^-)$. The $O(\sqrt{N})$ finite size correction is thus zero. In the ferromagnetic phase instead, the messages $u^+$ and $u^-$ become non-trivially correlated.

We computed with the population dynamics algorithm the solution of the fixed point Eq. \eqref{puu} at temperature $T=0$, for many values of the coupling $J$ and for connectivity $z=4$. The expectations we find in the expression of $\Delta^2$ given Eq. \eqref{deltarrg} are then computed sampling from the population. 

As in the case of the fully connected model, it is easier to verify the analytical predictions working at zero temperature, since the free energy and the energy coincides and the ground state of the system can be obtained through exact polynomial algorithms.

\begin{figure}
\includegraphics[width=\columnwidth]{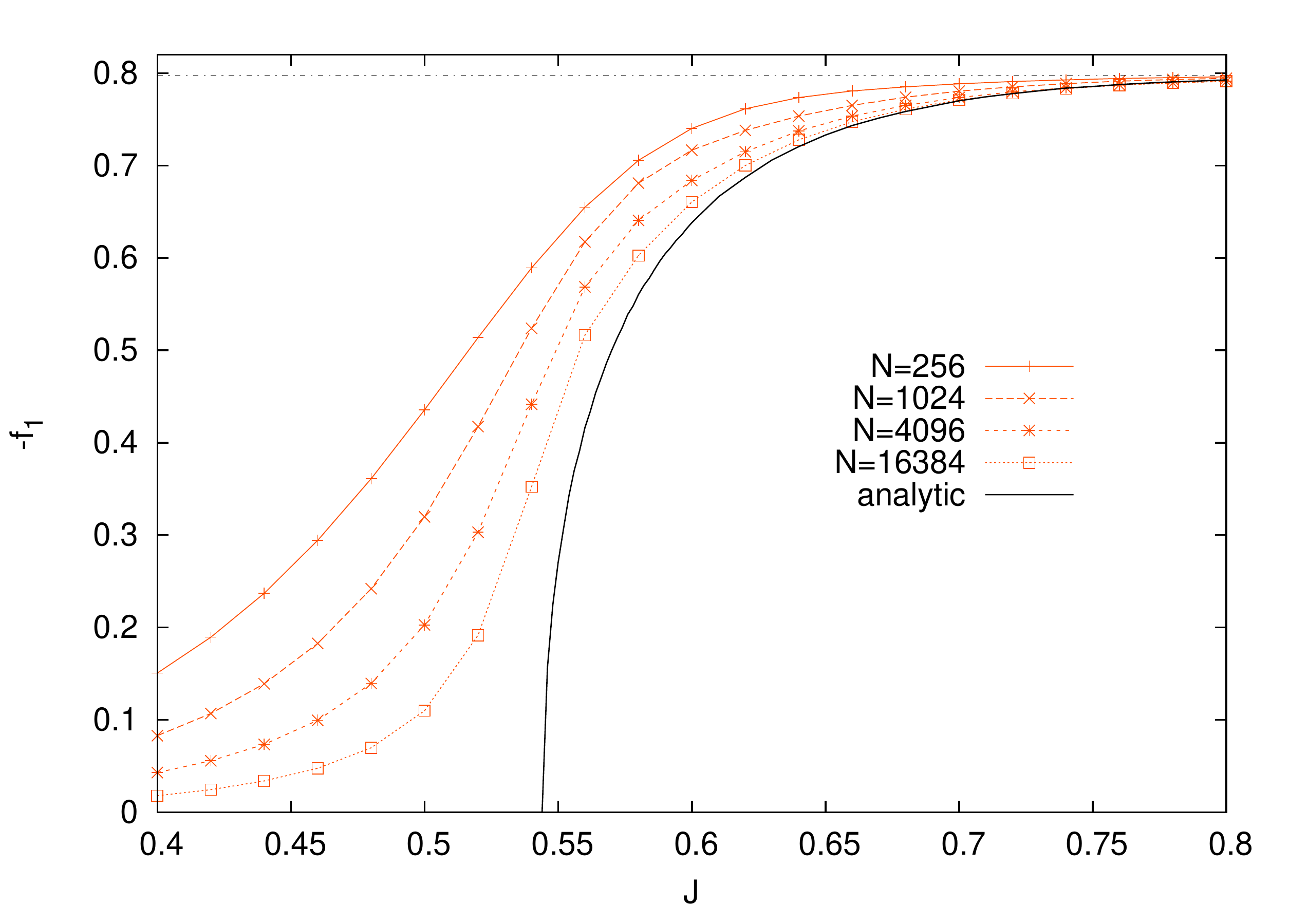}
\caption{The numerical estimates, obtained through the min-cut algorithm, of the coefficient of the $O(\sqrt{N})$ correction to the average free energy in the RFIM at zero temperature on Random Regular Graphs of connectivity $z=4$, for  many sizes of the systems (\textit{orange lines)}. The numerical curves (see Eq. \ref{f1tilde} for their definition)  extrapolate to the analytical prediction of Eqs. \eqref{f1delta} and \eqref{deltarrg} for large $N$ (\textit{black line}). The dashed black line is the asymptotic value $\sqrt{2/\pi}$.}
\label{fig:rrg}
\end{figure}

Therefore the analytical result is compared with an exact numerical algorithm that exploit the equivalence between the problem of finding the ground state of the RFIM on an arbitrary graph and  the minimum cut optimization problem \cite{Picard1975}. We used the implementation of the Goldberg-Tarjan's preflow push-relabel algorithm provided by the open source \text{Lemon Graph Library} \cite{Dezso2011}, whose worst case complexity is $\Theta(N^\frac{5}{2})$ for instances of the RRG ensemble.
The numerical estimate of the $O(\sqrt{N})$ coefficient is obtained from a linear combinations of the free energy of systems of different sizes, and it is given by
\begin{equation}
\tilde f_1(N) = \frac{\mathbb{E}[F(2N)-2F(N)]}{c \sqrt{N}}
\label{f1tilde}
\end{equation}
with $c= \frac{1}{\sqrt{2}}-1$. The value of $\tilde f_1(N)$, obtained averaging the minimum cut results over many samples of the system, should converge for large $N$ to the analytical value computed trough the population dynamic algorithm applied to Eq. \eqref{deltarrg}. The data plotted in Fig. \ref{fig:rrg} show a very good agreement between experiments and predictions, although the convergence is slow due to the presence of subleading $O(\frac{1}{\sqrt{N}})$ finite size effects.

\begin{figure}
\includegraphics[width=\columnwidth]{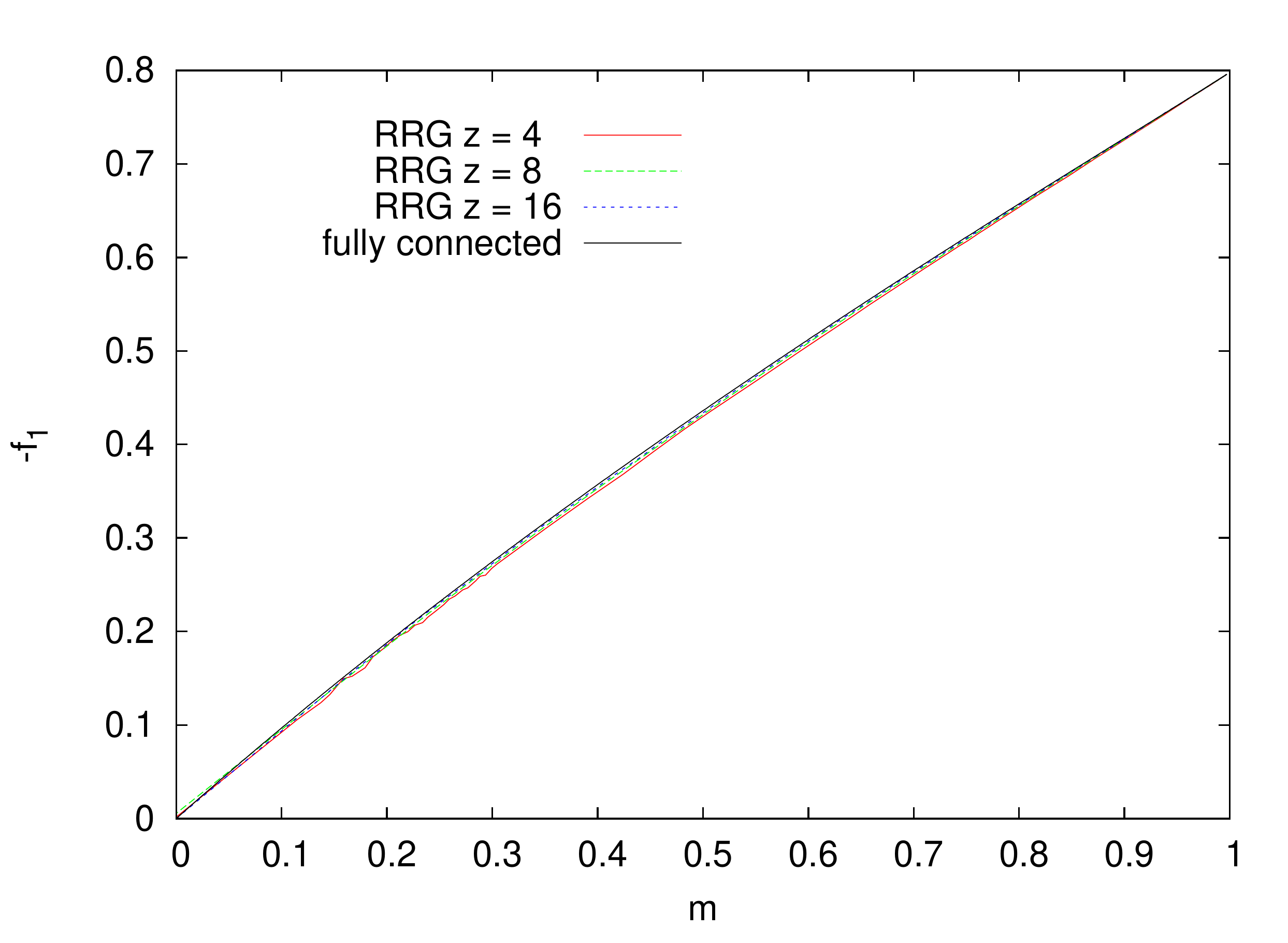}
\caption{The coefficient of the $O(\sqrt{N})$ correction to the average free energy versus the average absolute magnetization $m$ in the RFIM at zero temperature on Random Regular Graphs of different connectivities, as computed solving Eqs.  \eqref{puu} and  \eqref{deltarrg} with a population dynamic algorithm. The black line shows the same quantity for the Fully Connected model.}
\label{fig:rrg-pd}
\end{figure}

Moreover the analytic value of $f_1$, when plotted as a function of the absolute magnetization of the system, $m \equiv \lim_{N\to\infty} N^{-1}\mathbb E[ |\sum_i <\s_i> |]$, see Figure \ref{fig:rrg-pd}, shows a weak dependence from the connectivity of the system and an almost linear behaviour. A simple estimate of $f_1$ on both the fully connected graph and RRGs is then given by $f_1 \approx -\sqrt{2/ \pi}\  m$. 

\section{Conclusions}
\label{sec:concl}
We are arguing for the existence of an anomalous $O(1/\sqrt{N})$ subleading correction to the thermodynamic average free energy $f_0$ of systems with a zero-mean external random field. This correction is limited to the ferromagnetic phase, since it is caused by the difference of free energy among the two pure states.  If one is interested only in asymptotic  quantities such as $f_0$ , the slowing down of the convergence can be avoided in numerical simulations such as Markov Chain Monte Carlo methods choosing an initial condition uncorrelated to the realization of the disorder (as it is indeed usually done), such that the dynamics gets trapped  with equal probability in the state with lowest free energy or in the other one. On the other hand estimations of some observables on systems of finite size through exact algorithms, such as the minimum cut algorithm, are bound to follow a behaviour of the type $a(N)\sim a_0 + \frac{a_1}{\sqrt{N}}$, therefore the convergence is much slower than the usual $O(\frac{1}{N})$ behaviour one finds in the paramagnetic phase or in pure systems. Constraining the sampling an observable to a state symmetrically fluctuating around its mean asymptotic value is in this sense more convenient than sampling from the pure state with greater statistical weight.

Using a formalism inspired by some recent works \cite{Aspelmeier2003a, Parisi2008, Parisi2009, Parisi2010a} we presented a general framework to compute the coefficient $f_1$ of the $O(\sqrt{N})$ term in the average free energy. The computation has been carried out, using a variant of the replica trick, in two solvable mean field systems, the RFIM on the fully connected graph and on the RRG ensemble. The analytical results obtained, Eqs. \eqref{f1delta}\eqref{deltaFC}\eqref{deltarrg}, are found to be in strong agreement with the numerical simulations.

A different but equivalent approach to the problem, based on a peculiar replica symmetry breaking scheme, can be taken using the techniques of Refs. \cite{Dotsenko1997,Dotsenko2006}. Instead of computing the large deviation function $\phi(n,m)$, where $n$ and $m$ replicas are constrained to be in the up and down state respectively, and expanding to small values of $n$ and $m$, the same conclusions can be obtained summing over all the saddle point contributions obtained from the partitioning of a the replicas in the two sets. We preferred the approach of Refs. \cite{Aspelmeier2003a,Parisi2008, Parisi2009, Parisi2010a} because it is conceptually more clear and analytically less involving (although they share many similarities).

While the formalism we have developed in Section \ref{sec:formalism} is completely general, the exact computation of the coefficient $\Delta^2$ (thus of $f_1$) can be achieved only in mean field models. In finite dimension one has to resort to a perturbative diagrammatic expansion of a replicated field theory. We did not take this path, but our numerical simulations with a min-cut algorithm, using the same procedure described in Section \ref{sec:rrg} for the RRG ensemble, show, qualitatively and also quantitatively, the scenario depicted in Fig. \ref{fig:rrg}. Summing it up, as a general feature of models with zero-mean random external field, the average free energy density as a first finite size correction of order $O(\frac{1}{N})$ in the paramagnetic phase, and of order $O(\frac{1}{\sqrt{N}})$ in the ferromagnetic phase.

The research leading to these results has received funding from the European Research Council under the European Union's Seventh Framework Programme (FP7/2007-2013) / ERC grant agreement No. 247328 and from the
Italian Research Minister through the FIRB project No. RBFR086NN1.

\bibliography{bibliography}

\begin{thebibliography}{24}%
\makeatletter
\providecommand \@ifxundefined [1]{%
 \@ifx{#1\undefined}
}%
\providecommand \@ifnum [1]{%
 \ifnum #1\expandafter \@firstoftwo
 \else \expandafter \@secondoftwo
 \fi
}%
\providecommand \@ifx [1]{%
 \ifx #1\expandafter \@firstoftwo
 \else \expandafter \@secondoftwo
 \fi
}%
\providecommand \natexlab [1]{#1}%
\providecommand \enquote  [1]{``#1''}%
\providecommand \bibnamefont  [1]{#1}%
\providecommand \bibfnamefont [1]{#1}%
\providecommand \citenamefont [1]{#1}%
\providecommand \href@noop [0]{\@secondoftwo}%
\providecommand \href [0]{\begingroup \@sanitize@url \@href}%
\providecommand \@href[1]{\@@startlink{#1}\@@href}%
\providecommand \@@href[1]{\endgroup#1\@@endlink}%
\providecommand \@sanitize@url [0]{\catcode `\\12\catcode `\$12\catcode
  `\&12\catcode `\#12\catcode `\^12\catcode `\_12\catcode `\%12\relax}%
\providecommand \@@startlink[1]{}%
\providecommand \@@endlink[0]{}%
\providecommand \url  [0]{\begingroup\@sanitize@url \@url }%
\providecommand \@url [1]{\endgroup\@href {#1}{\urlprefix }}%
\providecommand \urlprefix  [0]{URL }%
\providecommand \Eprint [0]{\href }%
\providecommand \doibase [0]{http://dx.doi.org/}%
\providecommand \selectlanguage [0]{\@gobble}%
\providecommand \bibinfo  [0]{\@secondoftwo}%
\providecommand \bibfield  [0]{\@secondoftwo}%
\providecommand \translation [1]{[#1]}%
\providecommand \BibitemOpen [0]{}%
\providecommand \bibitemStop [0]{}%
\providecommand \bibitemNoStop [0]{.\EOS\space}%
\providecommand \EOS [0]{\spacefactor3000\relax}%
\providecommand \BibitemShut  [1]{\csname bibitem#1\endcsname}%
\let\auto@bib@innerbib\@empty
\bibitem [{\citenamefont {Fishman}\ and\ \citenamefont
  {Aharony}(1979)}]{Fishman1979}%
  \BibitemOpen
  \bibfield  {author} {\bibinfo {author} {\bibfnamefont {S.}~\bibnamefont
  {Fishman}}\ and\ \bibinfo {author} {\bibfnamefont {A.}~\bibnamefont
  {Aharony}},\ }\href {\doibase 10.1088/0022-3719/12/18/006} {\bibfield
  {journal} {\bibinfo  {journal} {Journal of Physics C: Solid State Physics}\
  }\textbf {\bibinfo {volume} {12}},\ \bibinfo {pages} {L729} (\bibinfo {year}
  {1979})}\BibitemShut {NoStop}%
\bibitem [{\citenamefont {Dagotto}(2005)}]{Dagotto2005}%
  \BibitemOpen
  \bibfield  {author} {\bibinfo {author} {\bibfnamefont {E.}~\bibnamefont
  {Dagotto}},\ }\href {\doibase 10.1126/science.1107559} {\bibfield  {journal}
  {\bibinfo  {journal} {Science}\ }\textbf {\bibinfo {volume} {309}},\ \bibinfo
  {pages} {257} (\bibinfo {year} {2005})}\BibitemShut {NoStop}%
\bibitem [{\citenamefont {Vink}\ \emph {et~al.}(2006)\citenamefont {Vink},
  \citenamefont {Binder},\ and\ \citenamefont {L\"{o}wen}}]{Vink2006}%
  \BibitemOpen
  \bibfield  {author} {\bibinfo {author} {\bibfnamefont {R.}~\bibnamefont
  {Vink}}, \bibinfo {author} {\bibfnamefont {K.}~\bibnamefont {Binder}}, \ and\
  \bibinfo {author} {\bibfnamefont {H.}~\bibnamefont {L\"{o}wen}},\ }\href
  {\doibase 10.1103/PhysRevLett.97.230603} {\bibfield  {journal} {\bibinfo
  {journal} {Physical Review Letters}\ }\textbf {\bibinfo {volume} {97}},\
  \bibinfo {pages} {230603} (\bibinfo {year} {2006})}\BibitemShut {NoStop}%
\bibitem [{\citenamefont {Krzakala}\ \emph {et~al.}(2011)\citenamefont
  {Krzakala}, \citenamefont {Ricci-Tersenghi}, \citenamefont {Sherrington},\
  and\ \citenamefont {Zdeborov\'{a}}}]{Krzakala2010}%
  \BibitemOpen
  \bibfield  {author} {\bibinfo {author} {\bibfnamefont {F.}~\bibnamefont
  {Krzakala}}, \bibinfo {author} {\bibfnamefont {F.}~\bibnamefont
  {Ricci-Tersenghi}}, \bibinfo {author} {\bibfnamefont {D.}~\bibnamefont
  {Sherrington}}, \ and\ \bibinfo {author} {\bibfnamefont {L.}~\bibnamefont
  {Zdeborov\'{a}}},\ }\href {\doibase 10.1088/1751-8113/44/4/042003} {\bibfield
   {journal} {\bibinfo  {journal} {Journal of Physics A: Mathematical and
  Theoretical}\ }\textbf {\bibinfo {volume} {44}},\ \bibinfo {pages} {042003}
  (\bibinfo {year} {2011})},\ \Eprint {http://arxiv.org/abs/1008.4497}
  {arXiv:1008.4497} \BibitemShut {NoStop}%
\bibitem [{\citenamefont {Krzakala}\ \emph {et~al.}(2010)\citenamefont
  {Krzakala}, \citenamefont {Ricci-Tersenghi},\ and\ \citenamefont
  {Zdeborov\'{a}}}]{Krzakala2010a}%
  \BibitemOpen
  \bibfield  {author} {\bibinfo {author} {\bibfnamefont {F.}~\bibnamefont
  {Krzakala}}, \bibinfo {author} {\bibfnamefont {F.}~\bibnamefont
  {Ricci-Tersenghi}}, \ and\ \bibinfo {author} {\bibfnamefont {L.}~\bibnamefont
  {Zdeborov\'{a}}},\ }\href {\doibase 10.1103/PhysRevLett.104.207208}
  {\bibfield  {journal} {\bibinfo  {journal} {Physical Review Letters}\
  }\textbf {\bibinfo {volume} {104}},\ \bibinfo {pages} {207208} (\bibinfo
  {year} {2010})},\ \Eprint {http://arxiv.org/abs/0911.1551} {arXiv:0911.1551}
  \BibitemShut {NoStop}%
\bibitem [{\citenamefont {Fytas}\ and\ \citenamefont
  {Martin-Mayor}(2013)}]{Fytas2013}%
  \BibitemOpen
  \bibfield  {author} {\bibinfo {author} {\bibfnamefont {N.~G.}\ \bibnamefont
  {Fytas}}\ and\ \bibinfo {author} {\bibfnamefont {V.}~\bibnamefont
  {Martin-Mayor}},\ }\href {\doibase 10.1103/PhysRevLett.110.227201} {\bibfield
   {journal} {\bibinfo  {journal} {Physical Review Letters}\ }\textbf {\bibinfo
  {volume} {110}},\ \bibinfo {pages} {227201} (\bibinfo {year}
  {2013})}\BibitemShut {NoStop}%
\bibitem [{\citenamefont {Ferrari}\ \emph {et~al.}(2013)\citenamefont
  {Ferrari}, \citenamefont {Lucibello}, \citenamefont {Morone}, \citenamefont
  {Parisi}, \citenamefont {Ricci-Tersenghi},\ and\ \citenamefont
  {Rizzo}}]{LucibelloER}%
  \BibitemOpen
  \bibfield  {author} {\bibinfo {author} {\bibfnamefont {U.}~\bibnamefont
  {Ferrari}}, \bibinfo {author} {\bibfnamefont {C.}~\bibnamefont {Lucibello}},
  \bibinfo {author} {\bibfnamefont {F.}~\bibnamefont {Morone}}, \bibinfo
  {author} {\bibfnamefont {G.}~\bibnamefont {Parisi}}, \bibinfo {author}
  {\bibfnamefont {F.}~\bibnamefont {Ricci-Tersenghi}}, \ and\ \bibinfo {author}
  {\bibfnamefont {T.}~\bibnamefont {Rizzo}},\ }\href {\doibase
  10.1103/PhysRevB.88.184201} {\bibfield  {journal} {\bibinfo  {journal}
  {Physical Review B}\ }\textbf {\bibinfo {volume} {88}} (\bibinfo {year}
  {2013}),\ 10.1103/PhysRevB.88.184201}\BibitemShut {NoStop}%
\bibitem [{\citenamefont {Lucibello}\ \emph {et~al.}(2014)\citenamefont
  {Lucibello}, \citenamefont {Morone}, \citenamefont {Parisi}, \citenamefont
  {Ricci-Tersenghi},\ and\ \citenamefont {Rizzo}}]{LucibelloRRG}%
  \BibitemOpen
  \bibfield  {author} {\bibinfo {author} {\bibfnamefont {C.}~\bibnamefont
  {Lucibello}}, \bibinfo {author} {\bibfnamefont {F.}~\bibnamefont {Morone}},
  \bibinfo {author} {\bibfnamefont {G.}~\bibnamefont {Parisi}}, \bibinfo
  {author} {\bibfnamefont {F.}~\bibnamefont {Ricci-Tersenghi}}, \ and\ \bibinfo
  {author} {\bibfnamefont {T.}~\bibnamefont {Rizzo}},\ }\href
  {http://arxiv.org/abs/1403.6049} {\bibfield  {journal} {\bibinfo  {journal}
  {ArXiv e-prints}\ ,\ \bibinfo {pages} {1}} (\bibinfo {year} {2014})},\
  \Eprint {http://arxiv.org/abs/1403.6049} {arXiv:1403.6049} \BibitemShut
  {NoStop}%
\bibitem [{\citenamefont {Montanari}\ and\ \citenamefont
  {Rizzo}(2005)}]{Montanari2005}%
  \BibitemOpen
  \bibfield  {author} {\bibinfo {author} {\bibfnamefont {A.}~\bibnamefont
  {Montanari}}\ and\ \bibinfo {author} {\bibfnamefont {T.}~\bibnamefont
  {Rizzo}},\ }\href {\doibase 10.1088/1742-5468/2005/10/P10011} {\bibfield
  {journal} {\bibinfo  {journal} {Journal of Statistical Mechanics: Theory and
  Experiment}\ }\textbf {\bibinfo {volume} {2005}},\ \bibinfo {pages} {P10011}
  (\bibinfo {year} {2005})},\ \Eprint {http://arxiv.org/abs/0506769v1}
  {arXiv:0506769v1 [cond-mat]} \BibitemShut {NoStop}%
\bibitem [{\citenamefont {Chertkov}\ and\ \citenamefont
  {Chernyak}(2006)}]{Chertkov2006}%
  \BibitemOpen
  \bibfield  {author} {\bibinfo {author} {\bibfnamefont {M.}~\bibnamefont
  {Chertkov}}\ and\ \bibinfo {author} {\bibfnamefont {V.~Y.}\ \bibnamefont
  {Chernyak}},\ }\href {\doibase 10.1103/PhysRevE.73.065102} {\bibfield
  {journal} {\bibinfo  {journal} {Physical Review E}\ }\textbf {\bibinfo
  {volume} {73}} (\bibinfo {year} {2006}),\ 10.1103/PhysRevE.73.065102},\
  \Eprint {http://arxiv.org/abs/0601487v2} {arXiv:0601487v2 [cond-mat]}
  \BibitemShut {NoStop}%
\bibitem [{\citenamefont {Parisi}\ and\ \citenamefont
  {Slanina}(2006)}]{Parisi2006}%
  \BibitemOpen
  \bibfield  {author} {\bibinfo {author} {\bibfnamefont {G.}~\bibnamefont
  {Parisi}}\ and\ \bibinfo {author} {\bibfnamefont {F.}~\bibnamefont
  {Slanina}},\ }\href {\doibase 10.1088/1742-5468/2006/02/L02003} {\bibfield
  {journal} {\bibinfo  {journal} {Journal of Statistical Mechanics: Theory and
  Experiment}\ }\textbf {\bibinfo {volume} {2006}},\ \bibinfo {pages} {L02003}
  (\bibinfo {year} {2006})},\ \Eprint {http://arxiv.org/abs/0512529v1}
  {arXiv:0512529v1 [cond-mat]} \BibitemShut {NoStop}%
\bibitem [{\citenamefont {Dotsenko}\ and\ \citenamefont
  {M\'{e}zard}(1997)}]{Dotsenko1997}%
  \BibitemOpen
  \bibfield  {author} {\bibinfo {author} {\bibfnamefont {V.}~\bibnamefont
  {Dotsenko}}\ and\ \bibinfo {author} {\bibfnamefont {M.}~\bibnamefont
  {M\'{e}zard}},\ }\href {\doibase 10.1088/0305-4470/30/10/015} {\bibfield
  {journal} {\bibinfo  {journal} {Journal of Physics A: Mathematical and
  General}\ }\textbf {\bibinfo {volume} {30}},\ \bibinfo {pages} {3363}
  (\bibinfo {year} {1997})},\ \Eprint {http://arxiv.org/abs/9611017}
  {arXiv:9611017 [cond-mat]} \BibitemShut {NoStop}%
\bibitem [{\citenamefont {Dotsenko}(2006)}]{Dotsenko2006}%
  \BibitemOpen
  \bibfield  {author} {\bibinfo {author} {\bibfnamefont {V.}~\bibnamefont
  {Dotsenko}},\ }\href {\doibase 10.1088/1742-5468/2006/06/P06003} {\bibfield
  {journal} {\bibinfo  {journal} {Journal of Statistical Mechanics: Theory and
  Experiment}\ }\textbf {\bibinfo {volume} {2006}},\ \bibinfo {pages} {P06003}
  (\bibinfo {year} {2006})},\ \Eprint {http://arxiv.org/abs/0603713}
  {arXiv:0603713 [cond-mat]} \BibitemShut {NoStop}%
\bibitem [{\citenamefont {Aspelmeier}\ \emph {et~al.}(2003)\citenamefont
  {Aspelmeier}, \citenamefont {Moore},\ and\ \citenamefont
  {Young}}]{Aspelmeier2003a}%
  \BibitemOpen
  \bibfield  {author} {\bibinfo {author} {\bibfnamefont {T.}~\bibnamefont
  {Aspelmeier}}, \bibinfo {author} {\bibfnamefont {M.~A.}\ \bibnamefont
  {Moore}}, \ and\ \bibinfo {author} {\bibfnamefont {A.~P.}\ \bibnamefont
  {Young}},\ }\href {\doibase 10.1103/PhysRevLett.90.127202} {\bibfield
  {journal} {\bibinfo  {journal} {Physical Review Letters}\ }\textbf {\bibinfo
  {volume} {90}},\ \bibinfo {pages} {127202} (\bibinfo {year}
  {2003})}\BibitemShut {NoStop}%
\bibitem [{\citenamefont {Parisi}\ and\ \citenamefont
  {Rizzo}(2008)}]{Parisi2008}%
  \BibitemOpen
  \bibfield  {author} {\bibinfo {author} {\bibfnamefont {G.}~\bibnamefont
  {Parisi}}\ and\ \bibinfo {author} {\bibfnamefont {T.}~\bibnamefont {Rizzo}},\
  }\href {\doibase 10.1103/PhysRevLett.101.117205} {\bibfield  {journal}
  {\bibinfo  {journal} {Physical Review Letters}\ }\textbf {\bibinfo {volume}
  {101}} (\bibinfo {year} {2008}),\ 10.1103/PhysRevLett.101.117205},\ \Eprint
  {http://arxiv.org/abs/0706.1180} {arXiv:0706.1180} \BibitemShut {NoStop}%
\bibitem [{\citenamefont {Parisi}\ and\ \citenamefont
  {Rizzo}(2009)}]{Parisi2009}%
  \BibitemOpen
  \bibfield  {author} {\bibinfo {author} {\bibfnamefont {G.}~\bibnamefont
  {Parisi}}\ and\ \bibinfo {author} {\bibfnamefont {T.}~\bibnamefont {Rizzo}},\
  }\href {\doibase 10.1103/PhysRevB.79.134205} {\bibfield  {journal} {\bibinfo
  {journal} {Physical Review B}\ }\textbf {\bibinfo {volume} {79}} (\bibinfo
  {year} {2009}),\ 10.1103/PhysRevB.79.134205},\ \Eprint
  {http://arxiv.org/abs/0811.1524v2} {arXiv:0811.1524v2} \BibitemShut {NoStop}%
\bibitem [{\citenamefont {Parisi}\ and\ \citenamefont
  {Rizzo}(2010)}]{Parisi2010a}%
  \BibitemOpen
  \bibfield  {author} {\bibinfo {author} {\bibfnamefont {G.}~\bibnamefont
  {Parisi}}\ and\ \bibinfo {author} {\bibfnamefont {T.}~\bibnamefont {Rizzo}},\
  }\href {\doibase 10.1088/1751-8113/43/4/045001} {\bibfield  {journal}
  {\bibinfo  {journal} {Journal of Physics A: Mathematical and Theoretical}\
  }\textbf {\bibinfo {volume} {43}} (\bibinfo {year} {2010}),\
  10.1088/1751-8113/43/4/045001},\ \Eprint {http://arxiv.org/abs/0910.4553}
  {arXiv:0910.4553} \BibitemShut {NoStop}%
\bibitem [{\citenamefont {Parisi}\ \emph {et~al.}(1987)\citenamefont {Parisi},
  \citenamefont {M\'{e}zard},\ and\ \citenamefont {Virasoro}}]{Parisi1987}%
  \BibitemOpen
  \bibfield  {author} {\bibinfo {author} {\bibfnamefont {G.}~\bibnamefont
  {Parisi}}, \bibinfo {author} {\bibfnamefont {M.}~\bibnamefont {M\'{e}zard}},
  \ and\ \bibinfo {author} {\bibfnamefont {M.~A.}\ \bibnamefont {Virasoro}},\
  }\href@noop {} {\emph {\bibinfo {title} {{Spin glass theory and beyond}}}}\
  (\bibinfo  {publisher} {World Scientific Singapore},\ \bibinfo {year}
  {1987})\BibitemShut {NoStop}%
\bibitem [{\citenamefont {Bray}\ and\ \citenamefont {Moore}(1985)}]{Bray1985}%
  \BibitemOpen
  \bibfield  {author} {\bibinfo {author} {\bibfnamefont {A.~J.}\ \bibnamefont
  {Bray}}\ and\ \bibinfo {author} {\bibfnamefont {M.~A.}\ \bibnamefont
  {Moore}},\ }\href {\doibase 10.1088/0022-3719/18/28/006} {\bibfield
  {journal} {\bibinfo  {journal} {Journal of Physics C: Solid State Physics}\
  }\textbf {\bibinfo {volume} {18}},\ \bibinfo {pages} {L927} (\bibinfo {year}
  {1985})}\BibitemShut {NoStop}%
\bibitem [{\citenamefont {Montanari}\ and\ \citenamefont
  {M\'{e}zard}(2009)}]{Montanari2009}%
  \BibitemOpen
  \bibfield  {author} {\bibinfo {author} {\bibfnamefont {A.}~\bibnamefont
  {Montanari}}\ and\ \bibinfo {author} {\bibfnamefont {M.}~\bibnamefont
  {M\'{e}zard}},\ }\href@noop {} {\emph {\bibinfo {title} {{Information,
  Physics and Computation}}}}\ (\bibinfo  {publisher} {Oxford Univ. Press},\
  \bibinfo {year} {2009})\BibitemShut {NoStop}%
\bibitem [{\citenamefont {Goldschmidt}\ and\ \citenamefont {{De
  Dominicis}}(1990)}]{DeDominicis1990}%
  \BibitemOpen
  \bibfield  {author} {\bibinfo {author} {\bibfnamefont {Y.~Y.}\ \bibnamefont
  {Goldschmidt}}\ and\ \bibinfo {author} {\bibfnamefont {C.}~\bibnamefont {{De
  Dominicis}}},\ }\href {\doibase 10.1103/PhysRevB.41.2184} {\bibfield
  {journal} {\bibinfo  {journal} {Physical Review B}\ }\textbf {\bibinfo
  {volume} {41}},\ \bibinfo {pages} {2184} (\bibinfo {year}
  {1990})}\BibitemShut {NoStop}%
\bibitem [{\citenamefont {Rizzo}\ \emph {et~al.}(2010)\citenamefont {Rizzo},
  \citenamefont {Lage-Castellanos}, \citenamefont {Mulet},\ and\ \citenamefont
  {Ricci-Tersenghi}}]{Rizzo2009}%
  \BibitemOpen
  \bibfield  {author} {\bibinfo {author} {\bibfnamefont {T.}~\bibnamefont
  {Rizzo}}, \bibinfo {author} {\bibfnamefont {A.}~\bibnamefont
  {Lage-Castellanos}}, \bibinfo {author} {\bibfnamefont {R.}~\bibnamefont
  {Mulet}}, \ and\ \bibinfo {author} {\bibfnamefont {F.}~\bibnamefont
  {Ricci-Tersenghi}},\ }\href {\doibase 10.1007/s10955-010-9938-3} {\bibfield
  {journal} {\bibinfo  {journal} {Journal of Statistical Physics}\ }\textbf
  {\bibinfo {volume} {139}},\ \bibinfo {pages} {375} (\bibinfo {year}
  {2010})},\ \Eprint {http://arxiv.org/abs/0906.2695} {arXiv:0906.2695}
  \BibitemShut {NoStop}%
\bibitem [{\citenamefont {Picard}\ and\ \citenamefont
  {Ratliff}(1975)}]{Picard1975}%
  \BibitemOpen
  \bibfield  {author} {\bibinfo {author} {\bibfnamefont {J.~C.}\ \bibnamefont
  {Picard}}\ and\ \bibinfo {author} {\bibfnamefont {H.~D.}\ \bibnamefont
  {Ratliff}},\ }\href {\doibase 10.1002/net.3230050405} {\bibfield  {journal}
  {\bibinfo  {journal} {Networks}\ }\textbf {\bibinfo {volume} {5}},\ \bibinfo
  {pages} {357} (\bibinfo {year} {1975})}\BibitemShut {NoStop}%
\bibitem [{\citenamefont {Dezső}\ \emph {et~al.}(2011)\citenamefont {Dezső},
  \citenamefont {J\"{u}ttner},\ and\ \citenamefont {Kov\'{a}cs}}]{Dezso2011}%
  \BibitemOpen
  \bibfield  {author} {\bibinfo {author} {\bibfnamefont {B.}~\bibnamefont
  {Dezső}}, \bibinfo {author} {\bibfnamefont {A.}~\bibnamefont {J\"{u}ttner}},
  \ and\ \bibinfo {author} {\bibfnamefont {P.}~\bibnamefont {Kov\'{a}cs}},\
  }\href {\doibase 10.1016/j.entcs.2011.06.003} {\bibfield  {journal} {\bibinfo
   {journal} {Electronic Notes in Theoretical Computer Science}\ }\textbf
  {\bibinfo {volume} {264}},\ \bibinfo {pages} {23} (\bibinfo {year}
  {2011})}\BibitemShut {NoStop}%
\end{thebibliography}%

\end{document}